\documentclass[prb,aps,twocolumn,epsfig,eqsecnum]{revtex4}
\usepackage{epsfig}
\usepackage{wasysym}
\usepackage{bm}
\usepackage{psfrag}
\usepackage{graphicx}
\usepackage{psfrag}
\newlength{\myL}

\newcommand{\be}{\begin{equation}}
\newcommand{\ee}{\end{equation}}
\newcommand{\bea}{\begin{eqnarray}}
\newcommand{\eea}{\end{eqnarray}}

\newcommand{\bB}{{\bf{B}}}
\newcommand{\bA}{{\bf{A}}}

\newcommand{\pref}[1]{(\ref{#1})}

\def\tit#1#2#3#4#5{{#1}{\bf #2}, #3 (#4)}

\def\npb{Nucl.\ Phys.\ B\ }

\def\prl{Phys.\ Rev.\ Lett.\ }

\def\prb{Phys.\ Rev.\ B\ }

\def\jsp{J.\ Stat.\ Phys.\ }

\def\mplb{Mod.\ Phys.\ Lett.\ B\ }

\def \r {{\bf r}}
\def \x {{\bf x}}
\def \r2 {\rho_2}
\def \r4 {\rho_4}
\def \r4t {\tilde{\rho}_4}

\begin{document}


\title{On bipartite Rokhsar-Kivelson points and Cantor deconfinement}

\author{Eduardo Fradkin,$^1$ David A. Huse,$^2$ R. Moessner,$^3$
V. Oganesyan$^2$ and S. L. Sondhi$^2$}

\affiliation{$^1$Department of Physics, University of Illinois at
Urbana-Champaign, Urbana, IL 61801, USA}

\affiliation{$^2$Department of Physics, Princeton University,
Princeton, NJ 08544, USA}

\affiliation{$^3$Laboratoire de Physique Th\'eorique
de l'Ecole Normale Sup\'erieure, CNRS-UMR8549, Paris, France}

\date{\today}

\begin{abstract}
Quantum dimer models on bipartite lattices exhibit Rokhsar-Kivelson
(RK) points with exactly known critical ground states and deconfined
spinons. We examine generic, weak, perturbations around these points.
In $d=2+1$ we find a first order transition between a ``plaquette''
valence bond crystal and a region with a devil's staircase of
commensurate and incommensurate valence bond crystals. In the
part of the phase diagram where the staircase is incomplete, the
incommensurate states exhibit a gapless photon and deconfined spinons
on a set of finite measure, almost but not quite a deconfined phase
in a compact $U(1)$ gauge theory in $d=2+1$! In $d=3+1$ we find
a continuous transition between the $U(1)$ resonating valence bond
(RVB) phase and a deconfined staggered valence bond crystal. In an
appendix we comment on analogous phenomena in quantum vertex models,
most notably the existence of a continuous transition on the triangular
lattice in $d=2+1$.
\end{abstract}


\maketitle

\section{Introduction}
\label{sec:intro}

Quantum dimer models (QDMs) were introduced by
Rokhsar and Kivelson\cite{Rokhsar88} to capture the low energy
dynamics of valence bond dominated phases of quantum Heisenberg
antiferromagnets.\cite{fn1} Work on these models has established that
their behavior on bipartite and non-bipartite lattices is
fundamentally different. On the latter, they admit RVB phases with
liquid dimer correlations and deconfined (test)
spinons\cite{MStrirvb,gregoirekag,msfishlat}
which have topological order of the $Z_2$/$BF$ variety.\cite{hos} On
the former they admit only crystalline
phases\cite{Rokhsar88,subirfinitesize,fradkiv,levitov,leung,ioffelar89,mcshex}
(valence bond crystals) in $d=2$, while supporting RVB phases of the
$U(1)$ or Coulomb variety in $d>2$.\cite{HKMS3ddimer,3dRVB,HermelePy}

One of the elegant features of the class of quantum dimer models
introduced in Ref.~\onlinecite{Rokhsar88}
is the existence of special points,
christened RK points, at which the exact ground state wavefunctions
are trivially determined and take the form of equal amplitude
superpositions over sets of states connected by the dynamics---the
classic RVB form. Additionally, test monomers have a vanishing
interaction at these points, which translates into the deconfinement
of gapped spinons in RVB language.\cite{MStrirvb}
The appropriate phase diagram for the square lattice case is sketched in 
Fig.~\ref{figure1}.

In this paper we study quantum dimer models
``near'' the RK points on bipartite lattices, {\it i.e.} QDMs that consist
of the RK point Hamiltonians plus small, but generic, perturbations.
The restriction to bipartite lattices brings an important simplification.
In a sense that we will make precise below, bipartite RK points are
described by gapless Gaussian field theories and this will greatly simplify
the task of analyzing perturbations.

Our basic new results are twofold. First, we establish the degree of
criticality of the RK points---more precisely of the associated Gaussian field
theories which are fixed points of the obvious renormalization group and to
which we shall refer for clarity as the RK fixed points (RKFPs). In $d=2+1$ we
show that on the honeycomb and square lattices the RKFPs describe
multicritical points with two relevant symmetric operators. In $d=3+1$, the
cubic lattice RKFP describes a critical point with just one relevant symmetric
operator.  Second, we show that the purely ``staggered'' phase that borders
the RK points in the simplest QDMs can be pushed a finite distance away in the
perturbed models and for $d=2+1$ the intermediate region is filled in by a
devil's staircase of commensurate and incommensurate valence bond crystals
which exhibit staggered order but also additional Bragg peaks. The
commensurate crystals are gapped and confining. The incommensurate crystals
exhibit a gapless photon (phason) and are deconfining, for sufficiently weak
perturbations. Further, although they occur at (boundary) points, they form a
generalized Cantor set of finite measure in the same limit. We refer to this
phenomenon, which comes remarkably close to a deconfined phase in a compact
$U(1)$ gauge theory in $d=2+1$, as Cantor deconfinement. In fact, near the RK
point this region is practically a deconfined phase, as the commensurate
confined phases occupy an asymptotically small fraction of the phase diagram
in this limit.  In addition, the gap in the confined phases is extremely
small.  The possibility of such phases was pointed out previously by
Levitov\cite{levitov} based on a mapping to the roughening problem.

This paper is organized as follows. In Section \ref{sec:height} we discuss the
height action for a $2+1$-dimensional (generalized) quantum dimer models on
the square and honeycomb lattices. In Section \ref{sec:honeycomb} we discuss
in detail the case of the honeycomb lattice and the mechanism that drives the
quantum phase transition first order in that case. In Section \ref{sec:tilted}
we discuss the development of the tilted phase and in Section \ref{sec:devil}
we show that due to the strong coupling nature of these generalized dimer
models, their tilted incommensurate phases generically exhibit a fully
developed devil's staircase analogous to the one discussed in two-dimensional
classical systems. In Section \ref{sec:square} we discuss the phase diagram
for the case of the square lattice, and in Section \ref{sec:landau} we discuss
the violation of the Landau rules in these quantum phase transitions, and the
deconfining nature of these quantum critical points, including the
relationship of RK points to the ``deconfined critical points'' discussed
recently by Senthil and collaborators.\cite{senthil2003} We summarize our
conclusions in Section \ref{sec:summary}. In the appendix, we provide a brief
discussion of analogous phenomena in quantum vertex models. Here, we find that
the RKFP in $d=2+1$ that controls the transition between a crystalline state
and a Cantor deconfined region is critical (rather than multicritical).

\section{Height action in $d=2+1$ dimensions}
\label{sec:height}

The Hamiltonian of the simplest
quantum dimer model introduced in Ref.~\onlinecite{Rokhsar88} has the
form:
\bea
\nonumber
H&=&-t\left
( |
\setlength{\unitlength}{3947sp}%
\begingroup\makeatletter\ifx\SetFigFont\undefined%
\gdef\SetFigFont#1#2#3#4#5{%
  \reset@font\fontsize{#1}{#2pt}%
  \fontfamily{#3}\fontseries{#4}\fontshape{#5}%
  \selectfont}%
\fi\endgroup%
\begin{picture}(155,154)(533,319)
\thicklines
\put(664,343){\circle{18}}
\put(556,342){\line( 1, 0){105}}
\put(557,343){\circle{18}}
\put(557,449){\circle{18}}
\put(556,449){\line( 1, 0){105}}
\put(664,449){\circle{18}}
\end{picture}
\rangle
\langle
\setlength{\unitlength}{3947sp}%
\begingroup\makeatletter\ifx\SetFigFont\undefined%
\gdef\SetFigFont#1#2#3#4#5{%
  \reset@font\fontsize{#1}{#2pt}%
  \fontfamily{#3}\fontseries{#4}\fontshape{#5}%
  \selectfont}%
\fi\endgroup%
\begin{picture}(154,155)(397,321)
\thicklines
\put(527,452){\circle{18}}
\put(528,344){\line( 0, 1){105}}
\put(527,345){\circle{18}}
\put(421,345){\circle{18}}
\put(421,344){\line( 0, 1){105}}
\put(421,452){\circle{18}}
\end{picture}
|+h.c.  \right) +v\left
( |
\setlength{\unitlength}{3947sp}%
\begingroup\makeatletter\ifx\SetFigFont\undefined%
\gdef\SetFigFont#1#2#3#4#5{%
  \reset@font\fontsize{#1}{#2pt}%
  \fontfamily{#3}\fontseries{#4}\fontshape{#5}%
  \selectfont}%
\fi\endgroup%
\begin{picture}(155,154)(533,319)
\thicklines
\put(664,343){\circle{18}}
\put(556,342){\line( 1, 0){105}}
\put(557,343){\circle{18}}
\put(557,449){\circle{18}}
\put(556,449){\line( 1, 0){105}}
\put(664,449){\circle{18}}
\end{picture}
\rangle \langle
\setlength{\unitlength}{3947sp}%
\begingroup\makeatletter\ifx\SetFigFont\undefined%
\gdef\SetFigFont#1#2#3#4#5{%
  \reset@font\fontsize{#1}{#2pt}%
  \fontfamily{#3}\fontseries{#4}\fontshape{#5}%
  \selectfont}%
\fi\endgroup%
\begin{picture}(155,154)(533,319)
\thicklines
\put(664,343){\circle{18}}
\put(556,342){\line( 1, 0){105}}
\put(557,343){\circle{18}}
\put(557,449){\circle{18}}
\put(556,449){\line( 1, 0){105}}
\put(664,449){\circle{18}}
\end{picture}
|+
|
\setlength{\unitlength}{3947sp}%
\begingroup\makeatletter\ifx\SetFigFont\undefined%
\gdef\SetFigFont#1#2#3#4#5{%
  \reset@font\fontsize{#1}{#2pt}%
  \fontfamily{#3}\fontseries{#4}\fontshape{#5}%
  \selectfont}%
\fi\endgroup%
\begin{picture}(154,155)(397,321)
\thicklines
\put(527,452){\circle{18}}
\put(528,344){\line( 0, 1){105}}
\put(527,345){\circle{18}}
\put(421,345){\circle{18}}
\put(421,344){\line( 0, 1){105}}
\put(421,452){\circle{18}}
\end{picture}
\rangle \langle
\setlength{\unitlength}{3947sp}%
\begingroup\makeatletter\ifx\SetFigFont\undefined%
\gdef\SetFigFont#1#2#3#4#5{%
  \reset@font\fontsize{#1}{#2pt}%
  \fontfamily{#3}\fontseries{#4}\fontshape{#5}%
  \selectfont}%
\fi\endgroup%
\begin{picture}(154,155)(397,321)
\thicklines
\put(527,452){\circle{18}}
\put(528,344){\line( 0, 1){105}}
\put(527,345){\circle{18}}
\put(421,345){\circle{18}}
\put(421,344){\line( 0, 1){105}}
\put(421,452){\circle{18}}
\end{picture}
|
\right)\\
H&=&-t\left( | \setlength{\unitlength}{3158sp}%
\begingroup\makeatletter\ifx\SetFigFont\undefined%
\gdef\SetFigFont#1#2#3#4#5{%
  \reset@font\fontsize{#1}{#2pt}%
  \fontfamily{#3}\fontseries{#4}\fontshape{#5}%
  \selectfont}%
\fi\endgroup%
\begin{picture}(194,185)(318,260)
\thicklines \put(468,421){\circle{18}} 
\put(358,420){\line( 1, 0){108}} 
\put(342,377){\circle{18}} \put(360,421){\circle{18}}
\put(396,284){\circle{18}} 
\multiput(434,282)(5.48824,9.14707){11}{\makebox(8.3333,12.5000){\SetFigFont{7}{8.4}{\rmdefault}{\mddefault}{\updefault}.}}
\multiput(340,378)(5.57647,-9.29412){11}{\makebox(8.3333,12.5000){\SetFigFont{7}{8.4}{\rmdefault}{\mddefault}{\updefault}.}}
\put(434,284){\circle{18}}
\put(488,377){\circle{18}}
\end{picture}
 \rangle
\langle
\setlength{\unitlength}{3158sp}%
\begingroup\makeatletter\ifx\SetFigFont\undefined%
\gdef\SetFigFont#1#2#3#4#5{%
  \reset@font\fontsize{#1}{#2pt}%
  \fontfamily{#3}\fontseries{#4}\fontshape{#5}%
  \selectfont}%
\fi\endgroup%
\begin{picture}(194,185)(318,93)
\thicklines \put(468,117){\circle{18}} 
\put(358,118){\line( 1, 0){108}} 
\put(342,161){\circle{18}} \put(360,117){\circle{18}}
\put(396,254){\circle{18}} 
\multiput(434,256)(5.48824,-9.14707){11}{\makebox(8.3333,12.5000){\SetFigFont{7}{8.4}{\rmdefault}{\mddefault}{\updefault}.}}
\multiput(340,160)(5.57647,9.29412){11}{\makebox(8.3333,12.5000){\SetFigFont{7}{8.4}{\rmdefault}{\mddefault}{\updefault}.}}
\put(434,254){\circle{18}}
\put(488,161){\circle{18}}
\end{picture}
|+h.c.  \right) +v\left
( |
\setlength{\unitlength}{3158sp}%
\begingroup\makeatletter\ifx\SetFigFont\undefined%
\gdef\SetFigFont#1#2#3#4#5{%
  \reset@font\fontsize{#1}{#2pt}%
  \fontfamily{#3}\fontseries{#4}\fontshape{#5}%
  \selectfont}%
\fi\endgroup%
\begin{picture}(194,185)(318,260)
\thicklines \put(468,421){\circle{18}} 
\put(358,420){\line( 1, 0){108}} 
\put(342,377){\circle{18}} \put(360,421){\circle{18}}
\put(396,284){\circle{18}} 
\multiput(434,282)(5.48824,9.14707){11}{\makebox(8.3333,12.5000){\SetFigFont{7}{8.4}{\rmdefault}{\mddefault}{\updefault}.}}
\multiput(340,378)(5.57647,-9.29412){11}{\makebox(8.3333,12.5000){\SetFigFont{7}{8.4}{\rmdefault}{\mddefault}{\updefault}.}}
\put(434,284){\circle{18}}
\put(488,377){\circle{18}}
\end{picture}
\rangle \langle
\setlength{\unitlength}{3158sp}%
\begingroup\makeatletter\ifx\SetFigFont\undefined%
\gdef\SetFigFont#1#2#3#4#5{%
  \reset@font\fontsize{#1}{#2pt}%
  \fontfamily{#3}\fontseries{#4}\fontshape{#5}%
  \selectfont}%
\fi\endgroup%
\begin{picture}(194,185)(318,260)
\thicklines \put(468,421){\circle{18}} 
\put(358,420){\line( 1, 0){108}} 
\put(342,377){\circle{18}} \put(360,421){\circle{18}}
\put(396,284){\circle{18}} 
\multiput(434,282)(5.48824,9.14707){11}{\makebox(8.3333,12.5000){\SetFigFont{7}{8.4}{\rmdefault}{\mddefault}{\updefault}.}}
\multiput(340,378)(5.57647,-9.29412){11}{\makebox(8.3333,12.5000){\SetFigFont{7}{8.4}{\rmdefault}{\mddefault}{\updefault}.}}
\put(434,284){\circle{18}}
\put(488,377){\circle{18}}
\end{picture}
|+
|
\setlength{\unitlength}{3158sp}%
\begingroup\makeatletter\ifx\SetFigFont\undefined%
\gdef\SetFigFont#1#2#3#4#5{%
  \reset@font\fontsize{#1}{#2pt}%
  \fontfamily{#3}\fontseries{#4}\fontshape{#5}%
  \selectfont}%
\fi\endgroup%
\begin{picture}(194,185)(318,93)
\thicklines \put(468,117){\circle{18}} 
\put(358,118){\line( 1, 0){108}} 
\put(342,161){\circle{18}} \put(360,117){\circle{18}}
\put(396,254){\circle{18}} 
\multiput(434,256)(5.48824,-9.14707){11}{\makebox(8.3333,12.5000){\SetFigFont{7}{8.4}{\rmdefault}{\mddefault}{\updefault}.}}
\multiput(340,160)(5.57647,9.29412){11}{\makebox(8.3333,12.5000){\SetFigFont{7}{8.4}{\rmdefault}{\mddefault}{\updefault}.}}
\put(434,254){\circle{18}}
\put(488,161){\circle{18}}
\end{picture}
\rangle \langle
\setlength{\unitlength}{3158sp}%
\begingroup\makeatletter\ifx\SetFigFont\undefined%
\gdef\SetFigFont#1#2#3#4#5{%
  \reset@font\fontsize{#1}{#2pt}%
  \fontfamily{#3}\fontseries{#4}\fontshape{#5}%
  \selectfont}%
\fi\endgroup%
\begin{picture}(194,185)(318,93)
\thicklines \put(468,117){\circle{18}} 
\put(358,118){\line( 1, 0){108}} 
\put(342,161){\circle{18}} \put(360,117){\circle{18}}
\put(396,254){\circle{18}} 
\multiput(434,256)(5.48824,-9.14707){11}{\makebox(8.3333,12.5000){\SetFigFont{7}{8.4}{\rmdefault}{\mddefault}{\updefault}.}}
\multiput(340,160)(5.57647,9.29412){11}{\makebox(8.3333,12.5000){\SetFigFont{7}{8.4}{\rmdefault}{\mddefault}{\updefault}.}}
\put(434,254){\circle{18}}
\put(488,161){\circle{18}}
\end{picture}
|\right)\
\label{eq:Hqdm}
\eea
for the square and honeycomb lattices respectively.  Here, a sum over
all plaquettes of the lattice is implicit.
On the honeycomb
lattice,\cite{readsachsun,mcshex} and with less confidence on the
square lattice,\cite{Rokhsar88,subirfinitesize,ioffelar89,fradkiv,levitov,leung}
it is known that the
model exhibits three phases.

\begin{widetext}

\begin{figure}[h!]
\psfrag{C}{\large{columnar}}
\psfrag{P}{\large{plaquette}}
\psfrag{I}{\large{incommensurate}}
\psfrag{T}{\large{tilt}}
\psfrag{m}{max}
\psfrag{R}{\large{RK point}}
\psfrag{0}{$0$}
\psfrag{inf}{$-\infty$}
\psfrag{1}{$1$}
\psfrag{pinf}{$\infty$}
\psfrag{s}{\large{staggered}}
\psfrag{v}{$v/t$}
\begin{center}
\includegraphics[width=0.8\textwidth]{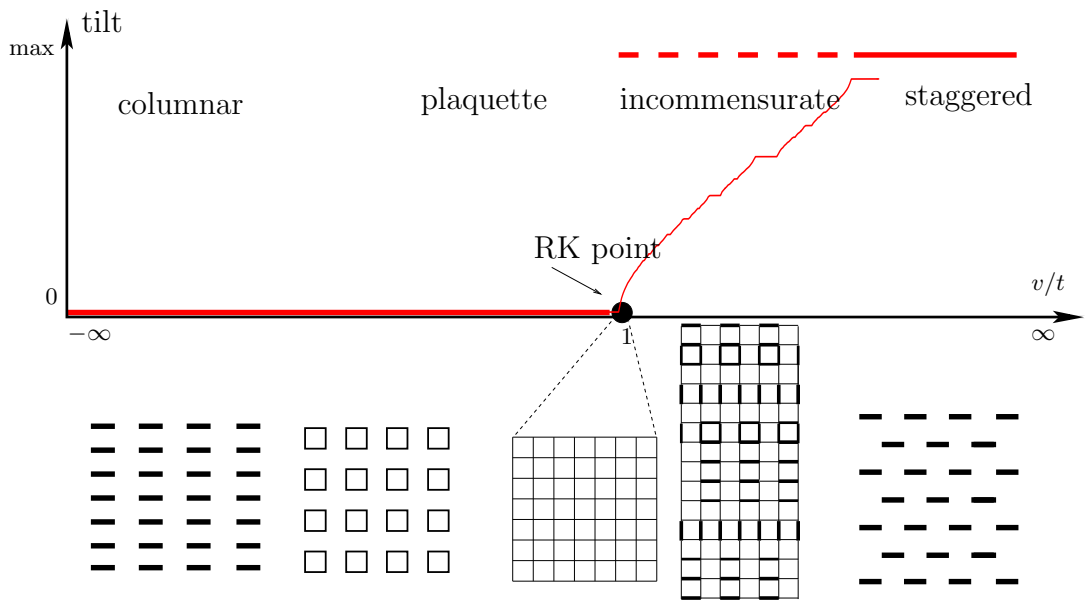}
\end{center}
\caption{
  Phase diagram of the square lattice quantum dimer model.  In the top part of
  the figure, the mean tilt of the height surface is plotted as a function of
  $v/t$ (see Section II). The corresponding dimer phases are sketched on the
  bottom part.  The flat side (columnar and plaquette solids) terminate in the
  RK critical point ($v/t=1$), which has no dimer long-range order.  Here, for
  the RK quantum dimer model (dashed line), the tilt jumps
  discontinuously, corresponding to a first order transition into the
  staggerred solid. 
  Upon inclusion of longer-ranged interactions, 
  however, the tilt will ascend a devil's
  staircase from the flat region, through a succession of incommensurate and
  commensurate phases. This phenomenon,
  (and the resulting
  structure factor) are discussed in detail in the text.
}
\label{figure1}
\end{figure}

\end{widetext}

For $v \ll t$ it is in a ``columnar''
crystal phase which gives way by a first order transition to a
``plaquette'' phase as $v/t$ is increased. The plaquette phase is
characterized by an order parameter that vanishes continuously
at $v=t$, the RK point. At the RK point there are multiple
ground states---the exact
ground state wavefunctions are the equal amplitude superpositions of
states in each sector of dimer configurations connected by the
(off-diagonal) resonance term
in Eq.~\pref{eq:Hqdm}. It is generally assumed that these sectors
are defined by two winding numbers (see
Ref.~\onlinecite{fradkin-book} for details)
although
a proof of ergodicity within sectors exists only for the zero winding
sector.\cite{thurston} For $v > t$ the system enters a
``staggered'' phase where the ground states contain no flippable
plaquettes.  Cartoons of these three phases --
columnar, plaquette and staggereed -- for
the square and honeycomb lattices can be found in
Figs.~1 of Refs.~\onlinecite{leung} and \onlinecite{mcshex},
respectively. The square case is reproduced in Fig.~\ref{figure1}.

On bipartite lattices the hardcore dimer constraint can be converted into a
Gauss law, $\nabla \cdot {\bf E} =0$. This is accomplished by endowing the
links of the lattice with an orientation, so that they point from one
sublattice to the other. Identifying an empty link with an electric flux $-1$
a link occupied by a dimer with a flux $z-1$ (where $z$ is the coordination of
the lattice) one obtains a lattice electric field ${\bf E}$ with the desired
property.
In $d=2+1$ this can
be solved by writing ${\bf E} = \nabla \times h$ in terms of a scalar
height field $h$.\cite{youngaxe,fradkin-kivelson} Here,
the violation of
the hardcore dimer covering constraint, through the
presence of a monomer or overlapping dimers,
shows up as a vortex in the height field.

In terms of $h$, the properties of the above QDMs in the vicinity of their
RK points are reproduced by the imaginary time Lagrangian
\begin{widetext}
\bea
{\cal L} &=& {1 \over 2} (\partial_\tau h)^2 +
{1 \over 2} \rho_2 (\nabla h)^2 + {1 \over 2} \rho_4 (\nabla^2 h)^2
+ \lambda \cos(2 \pi h)
\label{eq:heightact}
\eea
with $-\rho_2 \propto (v/t) + 1$ changing sign precisely at the RK
points.  The last term keeps track of the discreteness of the
microscopic heights, as is usual in height representations of problems
in $d=2$, and we will have more to say on that later. The square
and honeycomb lattice problems differ in the relevant values of
$\rho_4$ which are $(\pi/32)^2$ (square lattice) and $(\pi/18)^2$
(honeycomb). 

To determine the operator content of these theories, one also needs to specify
the compactification radius of the height variables. This is determined by the
minimal uniform shift of the heights which leads to an identical dimer
configuration. For the case of the honeycomb and square lattices, the
compactification radius equals 3 and 4, respectively.

Finally, the details of the lattice structure enter in the relationship
between lattice quantities and operators in the field theory.
Specifically, long wavelength correlations of the dimer densities $n_i$
(where $i$ labels the dimer direction) on the honeycomb lattice can be
reconstructed by means of the leading operator identifications,
\bea
n_1-\frac{1}{3}&=&\frac{1}{3}\partial_x h+\frac{1}{2}[ \exp(2\pi i h/3)
\exp(4\pi i x/3) + {\rm h.c.}] \nonumber\\
n_2-\frac{1}{3}&=&\frac{1}{3}(-\frac{1}{2}\partial_x+\frac{\sqrt{3}}{2}
\partial_y) h+\frac{1}{2}[ \exp(2\pi i h/3)
\exp(4\pi i x/3+4\pi i/3)+ {\rm h.c.}]          \\
n_3-\frac{1}{3}&=&\frac{1}{3}(-\frac{1}{2}\partial_x-\frac{\sqrt{3}}{2}
\partial_y) h+\frac{1}{2}[ \exp(2\pi i h/3)
\exp(4\pi i x/3-4\pi i/3) + {\rm h.c.}]\nonumber .
\eea
In these expressions, $x$ and $y$ denote Cartesian coordinates, the
$x$ axis being perpendicular to one dimer direction. The unit of
length is given by the separation of two neighbouring parallel dimers.
The corresponding expressions on the square lattice are
\bea
n_x-\frac{1}{4}&=&\frac{1}{4}(-1)^{x+y}\partial_y h +
\frac{1}{2}[(-1)^{x} \exp(\pi i h/2)+ {\rm h.c.}]\\
n_y-\frac{1}{4}&=& \frac{1}{4}(-1)^{x+y+1}\partial_x h+
\frac{1}{2}[(-1)^{y} \exp(\pi i h /2+\pi i /2)+ {\rm h.c.}]\nonumber
\eea
\end{widetext}

The effective Lagrangian of Eq. \pref{eq:heightact} was also considered in
Ref. \onlinecite{grinstein} in the context of the theory of classical Lifshitz
(multi)critical points in three dimensions and, for this reason, this
effective theory has been dubbed the quantum Lifshitz model in Ref.
\onlinecite{aff}. This effective Lagrangian exhibits a three-dimensional
analog of the physics of two-dimensional classical critical isotropic systems.
Thus, the action defined by Eq.~\pref{eq:heightact} exhibits a line of fixed
points parametrized by $\rho_4$ with $\rho_2=0$ and $\lambda=0$, under the
natural momentum and frequency shell renormalization group with $z=2$.  The RK
points on both lattices flow under this renormalization group, to RKFPs on
this line with the above values of $\rho_4$.  This accounts for their critical
correlations and a mode spectrum $\omega \sim k^2$.  In the $U(1)$ gauge
theoretic interpretation of the dimer model, this mode is a photon with an
anomalously soft dispersion. The irrelevance of the cosine is the necessary
vanishing of the instanton effects that otherwise gap the spectrum of such
compact confining theories \cite{polyakov} and this also accounts for the
deconfined spinons. The height action further accounts for two other
non-trivial features of the RK points, namely that they have degenerate ground
states in all winding number sectors (the RKFP point action is insensitive to
gradients of heights) and that for the zero-winding states the equal time
height correlations are logarithmic, \bea \left<h({\bf x}, \tau) h({\bf 0},
  \tau)\right>= -\frac {1}{4\pi \sqrt{\rho_4}} \ln(\pi |{\bf x}|/a) \eea and
precisely those of the classical dimer problem.  This in turn allows us to fix
the values of $\rho_4$ quoted earlier as the classical dimer
problem is exaclty solvable. 
(This also clarifies that these values are really the fixed point
values.)

On both lattices, the action (Eq.~\ref{eq:heightact}) accounts for the
plaquette crystals for $v<t$ as flat states of the height variable with
different local fluctuations: when $\rho_2 > 0$, $\lambda$ is relevant,
i.e. it is dangerously irrelevant at the RKFPs. The selection of
plaquette states indicates that $\lambda >0$ 
for $v/t\alt 1$ on both lattices.
For $v > t$, $\rho_2 < 0$: the height tilts, and in the
absence of any restoring term in Eq.~\ref{eq:heightact}
it attains its maximum tilt consistent with the microscopic
constraints. This tilted phase translates into staggered dimer correlations
which break translational symmetry on the square lattice and only
rotational symmetry on the honeycomb.

The transition between the plaquette and staggered phases is somewhat
unusual. As $\rho_2$ goes smoothly through zero at the RK point, the
plaquette order vanishes {\em continuously} there. Further, as
$v \rightarrow t^-$, there are two divergent length
scales: $\xi \sim |v/t -1|^{-1/2}$ and
$\xi_c \sim {1 \over \sqrt{\lambda}} 
\xi^\theta$, with $\theta = 6, {5 \over 2}$ 
on the square and honeycomb lattices,
respectively. The latter arises from the dangerously irrelevant
cosine and is the length scale on which the plaquette order appears;
the value of $\theta$ follows from the dimension of the irrelevant
cosine at the fixed point.
However, the staggered phase emerges at full strength immediately
and the ground state energy density has a derivative discontinuity
at the RK point which indicates a first order transition in the
thermodynamic classification.

Let us comment briefly on the origins of the effective action of
Eq.~\pref{eq:heightact}.
The height representation for dimers was used by several
workers on this problem; the recognition that it can be extended to
capture the physics of the RK point is due to Henley and a more complete
account of the properties of the RK point and the transition about it
was given in Ref.~\onlinecite{fn1}.
Henley's basic observation was that the
Hamiltonian Eq.~\ref{eq:Hqdm} also governs the master equation for the
probability distribution of the classical dimer problems endowed with
the simplest plaquette flip dynamics.\cite{henleyjsp} In height
variables the classical dimer problems are in the rough phase and so
it is reasonable that the dynamics is captured at
long wavelengths by the Langevin equation
\bea
\partial_\tau h (\x) = - {\delta S[h] \over \delta h(\x)} + \zeta(\x,\tau)
\eea
where $S[h] \sim \int d^2 x(\nabla h)^2$ is the coarse grained entropy
functional of the classical dimer problem. For Gaussian white noise
distributed $\zeta$ the average over trajectories is weighted by the
Lagrangian,
\bea
{\mathcal L}=\left(\partial_\tau h + {\delta S[h] \over \delta h(\x)}\right)^2
\eea
which, upon dropping a total time derivative, is the action governing
the RK points. 

This construction shows that the RK point action has higher dimension
operators than one might have guessed on symmetry grounds alone. Indeed,
the action \pref{eq:heightact} doesn't distinguish our two lattices with
their different
symmetries. As advertised, we are interested in this paper in weak, but
otherwise generic perturbations of the Hamiltonians \pref{eq:Hqdm}
about their
RK points. Such generic perturbations of the RK point Hamiltonians will
give rise to all operators allowed by symmetry and so we shall include
them in the analysis, with the proviso that they enter with small
couplings. A second implication of considering perturbed models will be
that the values of the couplings in \pref{eq:heightact}, in particular
the marginal coupling $\rho_4$ will also be somewhat different from
their values quoted earlier---these differences will again be small.
In the following we will refer to this set of modifications as working
near the RKFPs without further explanation.

A final comment is in order on an unusual aspect of the RKFPs. These
fixed points, and indeed the entire fixed line, exhibit an enlarged
symmetry in which the height field is free to tilt with any magnitude
in any direction---which is then spontaneously broken by the ground
states. While this appears to be non-generic, such extra symmetries are
a feature of any purely Gaussian fixed point. For instance at and above
$d=4$, the Ising transition is controlled by a Gaussian fixed point
that is invariant under constant shifts of the scalar field. 
Much as in that case we work near zero field (magnetization), in our problem
we will work near zero tilt. In both cases this is the correct choice for a
sector of the space of couplings---the sector in which the transition is
continuous in mean field theory.

\section{Expanded action on honeycomb lattice}
\label{sec:honeycomb}

We turn now to the honeycomb lattice.
For honeycomb dimers, the heights live on
the triangular lattice, which naively indicates an action rotationally
invariant to fourth order in gradients. However, the microscopic
heights have a sublattice structure: they take values $0$,$1$ and $2$
{\it modulo} $3$ on the three sublattices of the triangular lattice.
Consequently, the symmetries of rotation by $\pi/3$ and inversion
also involve sending $h \rightarrow -h$. It follows then that
the allowed non-irrelevant operators at the RKFP
include the relevant cubic term,
\bea
{\mathcal L}_3= g_3 \
(\partial_x h) ( {1 \over 2} \partial_x h - {\sqrt{3} \over 2} \partial_y h)
({1 \over 2} \partial_x h  + {\sqrt{3} \over 2} \partial_y h)
\eea
in addition to the marginal coupling,
\bea
{\mathcal L}_4 =g_4 \left[\nabla h \cdot \nabla h \right]^2 .
\eea
The growth of $g_3$ in the infrared indicates that the criticality
of the RK point does not survive its inclusion---the form of
the interaction indicates that when $g_3$ is nonzero the system has
already entered the tilted phase, although with the tilt now locked
to the lattice. This leads us to conclude that the transition between the
flat plaquette phase and the tilted phase goes truly first order.

If we tune $g_3$ to zero, we can examine the stability of the
multicritical RK point.
In a renormalization group treatment, where we keep a cutoff in
space alone, the lowest non-trivial order flows of the two
interactions are
\bea
{d \lambda \over d t} &=& - \left( {\pi \over 2 \rho_4^{1/2}} -2 \right)
\lambda \nonumber \\
{d g_4 \over d t} &=& -{9 \over 4 \pi \rho_4^{3/2}} g_4^2 \ ,
\eea
whence both interactions are irrelevant for $g_4 >0$.\cite{grinstein}
As $g_4$ is marginally
irrelevant it produces logarithmic corrections and the
flows also renormalize $\rho_4$ thus parking the system
elsewhere on the fixed line. With these observations we see
that the multicritical behavior represented by the RK point
is stable on a surface of codimension 2, provided the
long wavelength $\rho_4 \le (\pi/4)^2$.
The latter condition holds for small perturbations of
Eq.~(\ref{eq:heightact}).

\section{ The Tilted Phase}
\label{sec:tilted}

Once $g_4$ is large enough, the tilt no longer jumps to its maximum value on
leaving the flat phase. When $g_3$ is nonzero, the tilt prefers to point along
one of the dimer directions or in between two dimer directions depending on
the sign of this coupling. A similar role is played by $\tilde{g}_4$ on the
square lattice (see below). Here the restriction to small tilts is crucial:
the states for $g_3 >0$ and $g_3 <0$ are {\it not} related by symmetry; in
particular, at large tilts there are only states with one sign of the tilt,
i.e. with the dimers dominantly oriented along the three bond directions. We
also note that on traversing the multicritical point with $g_3=0$, the same
orientational selection is now effected by the fifth order term, 
\bea
{\mathcal L}_5&=&g_5 \ ( \nabla h)^2 (\partial_x h) ( {1 \over 2} \partial_x h
- {\sqrt{3} \over 2} \partial_y h) ({1 \over 2} \partial_x h + {\sqrt{3}
  \over 2} \partial_y h) 
\nonumber\\
&& 
\eea 
which is thus dangerously irrelevant at the multicritical fixed point.

The properties of the tilted phase can be analyzed by expanding in
small fluctuations about the weakly tilted state (henceforth we assume
$g_3<0$, so that the tilt is along the x-axis, and that $|g_3|$ is small)
\begin{equation}
h({\bf{r}},\tau)={\bf C}\cdot{\bf r}+ \delta h({\bf r},\tau),\
{\bf C}=C\ \hat{\bf x}.
\end{equation}
The corresponding Goldstone action is given (to quadratic order in fluctuations) by
\begin{equation}
\delta{\cal L}=\frac{1}{2}(\partial_\tau \delta
h)^2+\frac{\rho_4}{2}(\nabla^2 \delta h)^2+ \frac{v_l^2}{2}(\partial_x
\delta h)^2+\frac{v_t^2}{2}(\partial_y \delta h)^2.
\end{equation}
Higher order terms (in $\delta h$) are irrelevant due to non-divergence of
fluctuations. Provided $|\rho_2|<g_3^2/g_4$ then  $v_t \simeq v_l $ and a
single correlation length can be defined,  
$\xi^{-1}=v_l/\sqrt{\rho_4}$, so that for 
sufficiently large momenta, $q>\xi^{-1}$, the spectrum is quantum
critical, while as $q\rightarrow 0$ the modes acquire a stiffness, their
dispersion is linear with longitudinal 
and transverse velocities given by $v_l$ and $v_t$.
The average
tilt of the height variable translates into new Bragg peaks
appearing in the structure factor for the dimer densities (as
deduced from Eqs. 2.3
and
2.4). They are of two types: first is the
\emph{commensurate} Bragg peak at the wavevector of the maximally
``staggered" state located at the origin for the honeycomb case
(and at $(\pi,\pi)$ in the case of the square lattice), the second
type are the \emph{incommensurate} peaks displaced from the
characteristic wavevectors of the columnar/plaquette pattern (e.g. $(4\pi/3,0)$ and $(\pi,0)$ on honeycomb and square lattices, respectively) by an amount
proportional to $C$. In the tilted regime the intensity of the commensurate peak varies as $C^2$, while that of the incommensurate peaks as $\xi^{
-\frac{\pi}{9\sqrt{\rho_4}}}$ (on the square lattice as $\xi^{
-\frac{\pi}{16\sqrt{\rho_4}}}$). Generally, as a result of quantum critical
fluctuations, the two peaks differ in intensity
parametrically. Both their intensities are 
$C^2$ for $\rho_4$ corresponding to the the purely short range 
Hamiltonian of Eq.~\ref{eq:Hqdm}. The existence of gapless 
modes has important
implications, one of them is that test monomers (spinons) interact
through a potential that generally grows only logarithmically with
distance, as opposed to the linearly growing \emph{confining}
potential encountered in commensurate valence bond solids.

\section{Incommensurate locked crystal phases: the Devil's staircase}
\label{sec:devil}

The actual state of the system involves an interplay between the propensity of
the system to establish a tilt discussed thus far and the discreteness of the
microscopic heights. As we will quantify a little later, the operators
encoding the discreteness are much more irrelevant (although dangerously so,
see below) at the RKFP than those responsible for the selection of the tilt
magnitude and direction. Therefore, near the RKFP the problem can be treated
in the order in which we have discussed it. Away from this limit one can worry
about more complex phase diagrams.

In order to treat height discreteness properly, we need to
generalize our treatment of the discreteness of the height in
Eq.~\pref{eq:heightact}. There we included a potential that favors
discrete flat configurations of the height field but in doing so
we ignored the possibility of the height locking into tilted
configurations. To identify the full set of locking potentials, we
note that the height and the lattice points together define a
three dimensional lattice with points $(h,{\bf x})$. A general
locking potential, in particular the one generated from the
microscopic constraints under renormalization, can be expanded in
the set of reciprocal lattice vectors $\{\bf G\}$ of this 
three-dimensional
lattice,
\begin{equation}
V_{\rm lock}(h,{\bf x})
 = \sum_{\{\bf G\}} V_{\bf G} \;  e^{i(G_h h + \bf G_{\bf x}
\cdot \bf x)}
\end{equation}
and thus each $\bf G$ defines a locking potential. Nicely enough,
the height-space lattice for the honeycomb dimer problem is the
simple cubic lattice with the $[111]$ direction measuring height,
so working out the set of locking potentials is particularly
simple.

As discussed in the previous section, in the tilted phase the mean-square
fluctuations of the height are finite.  As a consequence of this, whenever the
average tilt of the height surface is commensurate with the lattice, the
corresponding terms in the locking potential (above) are asymptotically
relevant.  At such points the ground state is a commensurate valence bond
crystal with a gapped spectrum.  At incommensurate tilts the outcome depends
upon the strength of $V_{\rm lock}$ relative to the remaining quantum
fluctuations. When the latter are strong, as is the case for small tilts, the
ground state at most tilts is an incommensurate valence bond crystal with a
Bragg peak at the incommensurate wavevector but with a gapless (phason)
spectrum. In the gauge theoretic interpretation of the dimer model, this
excitation is a photon and correspondingly test monomers are deconfined---in
that they experience the logarithmic interaction of free two-dimensional
electrodynamics.  If the locking potential is increased and/or the quantum
fluctuations decreased, we expect a version of Aubry's ``breaking of
analyticity''\cite{Aubry} transition beyond which the incommensurate ground
states are pinned, their low-lying excitations are localized though still
gapless, and the incommensurate ground states occupy a set of measure zero in
the phase diagram.

The weak locking regime exists, parametrically, at small tilts
near the RK point. For concreteness, consider approaching the
multicritical surface by tuning $g_3$ with $\rho_2$ set equal to
zero. The closer we get to the transition, the higher order the
commensuration that leads to locking, with $|{\bf G}|\sim 1/C$ or
larger.  This has various effects: first the coefficients of the
corresponding locking potentials generally decrease as one goes to
larger ${\bf G}$; second, the operators become increasingly
irrelevant at the Rokhsar-Kivelson fixed point; and third, the
length scale increases at the crossover from the quantum critical
regime where the locking potential is irrelevant to the tilted
regime where it becomes relevant.  These effects combine together
to make the renormalized strength of the locking potential, and
thus the size of the gap in each commensurate phase, as well as
the width of the parameter range in the phase diagram occupied by
the phase, all vanish at small tilts $C$ exponentially in $1/C^2$:
e.g., the gap
\begin{equation}
\Delta\sim C^{a/(\rho_4C^2)}~,
\end{equation}
As the correlation length
$\xi\propto 1/C$, this is equivalent 
to an ordering length that grows rather fantastically, 
\bea
\xi_c \sim \xi^{a^\prime \xi^2/\rho_4} 
\eea
where $a, a^\prime$ are positive constants.
Consequently, even the commensurate phases are
for all practical purposes gapless in this weak-tilt regime. Thus
when the tilting transition is either continuous or very weakly
first-order, there is a deconfined, weakly-ordered dimer-crystal
phase near the RK point where the ordering is generically
incommensurate! By contrast, the existence of the strong locking
regime will need to be determined by actual solution of the
relevant models---the existence of the fully tilted states is not
proof of strong locking, merely of the existence of a maximum tilt
as a consequence of the periodicity of the height variable.

Returning to the infinite volume and zero temperature phase
diagram, the commensurate and incommensurate states will be
interleaved in the classic manner of the devil's staircase studied
in a variety of commensurate-incommensurate problems.\cite{cl} At
weak locking, close to the RK point, the staircase is incomplete,
i.e. the deconfined incommensurate points will form a set of
finite measure as the control parameter, $\rho_2$ or $g_3$, is
varied. In fact, the fraction of the phase diagram occupied by the
incommensurate deconfined states approaches unity as the RK point
is approached. This is the phenomenon we have termed Cantor
deconfinement. Farther away from the RK point the staircase could
become complete, with the incommensurate points being a set of
zero measure, before the maximally tilted phase is reached.
Needless to say, this scenario assumes that no first order
transitions intervene.

Finally, a brief comment on Bragg peaks.
The actual details of
the ordering---periodic in the commensurate states and
quasiperiodic in the incommensurate states---will cause the second
peak to fragment into multiple and even a dense set of
higher-order Bragg peaks.

\section{Expanded action on the square lattice}
\label{sec:square}

On the square lattice the generic case requires that we (a) allow $\rho_4$ to
vary, (b) include the quadratic and strictly marginal term,
\bea
{\cal L}_m = {1 \over 2}  \tilde{\rho_4}
\left[(\partial_x^2 h)^2 + (\partial_y^2 h)^2 \right] \ ,
\eea
and (c) include the interactions
\bea
{\cal L}_{\rm  int} =
  g_{4} \left[ \nabla h \cdot \nabla h \right]^2
 + \tilde{g}_{4} \left[ (\partial_x h)^4 + (\partial_y h)^4 \right]^2 \ .
\eea
The impact of (a) and (b) together is that we are interested in flows
about a two dimensional surface of fixed points. Unlike on
the honeycomb lattice, there is no cubic invariant so that the
transition is continuous at the level of mean field theory for the
height action. However, there
are two tree-level marginal couplings whose fate needs to be
decided by fluctuations.  Let us focus
for now on the flow around the fixed line with $\tilde{\rho}_4=0$.

The lowest non-trivial order flows of interest are:
\bea
{d \lambda \over d t} &=&  - \left( {\pi \over 2 \rho_4^{1/2}} -2 \right)
\lambda \nonumber \\
{d g_4 \over d t} &=& -{9 \over 4 \pi (\rho_4)^{3/2}}
(g_4^2 + g_4 \tilde{g}_4 + {1 \over 4} \tilde{g}_4^2) \nonumber \\
{d \tilde{g}_4 \over d t} &=& -{9 \over 4 \pi (\rho_4)^{3/2}}
({2 \over 3} g_4 \tilde{g}_4 + {1 \over 2} \tilde{g}_4^2) \ .
\eea
Interestingly, the flows in the $(g_4, \tilde{g}_4)$ plane are attracted
to the origin only along the positive $g_4$ axis; generically
the flows run away to the region where the action is unstable
at quartic order in gradients. From this we conclude that it
is highly likely that the transition is driven first order
by fluctuations in the generic case. Now the RK point behavior
requires a further fine tuning, $\tilde{g}_4 =0$, and hence it
represents the behavior of a multicritical surface of codimension
two.
We find that the inclusion of a small $\tilde{\rho}_4$ does
not alter the topology of the flows and there remains only one
trajectory flowing into the fixed point at the origin.

For the runaway flows, the pattern of symmetry breaking is indicated
by the initial sign of $\tilde{g}_4$ and depending on that we
find four states with the tilt either aligned or at angle $\pi/4$ to the
lattice axes. All of them exhibit, necessarily, a modulation of the
dimer density at wavevector $(\pi,\pi)$. If  $\tilde{g}_4 =0$,
the sixth order term is now responsible for the orientational
selection.

The analysis of the tilted phase now requires identifying the
reciprocal lattice vectors of the diamond lattice, with its
$[100]$ direction measuring the height. The qualitative features
of the devil's staircase are exactly as in the honeycomb case.
Bragg peaks now appear at $(\pi,\pi)$ and near $(\pi,0)$ and
$(0,\pi)$.

\section{The Cubic lattice}
\label{sec:cubic}

We turn finally to the case of the $d=3$
cubic lattice.  Now the constraint is solved by introducing
a gauge field $\bA$, in terms of which the action is
of the form,\cite{3dRVB}
\bea
{\cal L}=
 {1 \over 2} (\partial_t \bA)^2 + {1 \over 2} \rho_2
(\nabla \times \bA)^2 + {1 \over 2} \rho_4(\nabla\times \nabla \times\bA)^2
 \ ,
\label{eq:actiongauge}
\eea
with $\rho_2=0$ again signaling the RK point. In gauge-invariant notation,
the quadratic terms become the squares of the electric and 
the corresponding magnetic field in $d=3+1$.

The discreteness of the microscopic fields is irrelevant in a finite
interval around the RK point and hence can be ignored in its
proximity.  The only non-irrelevant
term consistent with cubic symmetry missing from
Eq.~\ref{eq:actiongauge} is
\bea
\tilde{\rho}_4 \
\sum_i (\partial_i (\nabla \times \bA)_i)^2
\ ,
\eea
which is strictly marginal. Hence the critical behavior represented
by the RK point is (marginally) stable. The additional
coupling leads to rotationally non-invariant critical points.

For $\rho_2 <0$, we need to include the (dangerously) irrelevant
terms that control the magnitude of the ``magnetic field'',
$\bB = \nabla \times \bA$. These are $\bB^4$ and $B_x^4 + B_y^4
+ B_z^4$ and lead to staggered crystals (now with a modulation of the
dimer density at $(\pi,\pi,\pi)$) aligned parallel or intermediate
to the lattice axes.
The connected correlations of the dimer density exhibit distorted
dipolar correlations. This spectrum exhibits a gapless photon with
fluctuations of the fields about their ground state values, and
deconfined spinons. Altogether, one obtains critical points
separating a liquid of dimers from a crystal. At larger tilts
confinement and further symmetry breaking can be expected to set
in, but that is outside the validity of our analysis. 

We also note that other related problems, such as the two-dimer model on the
bipartite diamond lattice studied by Hermele {\it et al.}\cite{HermelePy} in
its incarnation as a pyrochlore Ising antiferromagnet with Ising dynamics,
will {\it mutatis mutandis} exhibit the same phase diagram.

\section{Landau rules and order parameter theories}
\label{sec:landau}

In Ref.~\onlinecite{fn1},
three of us commented that the RK point in the
simplest bipartite QDMs in $d=2+1$ sat at the transition
between two symmetry incompatible valence bond crystals
and attributed the
divergence of the correlation length at it to the deconfinement
of the gauge/height field at the RK point. Recently
Senthil {\it et al.\/} \cite{senthil2003} have considered a
similar scenario for the transition between a Neel state and
a valence bond crystal, although now with dynamical spinons,
and christened the intervening critical point a deconfined
critical point.

At issue in these discussions are two distinct but closely
related possibilities. First, the violation of the
Landau rule that a phase transition between two symmetry
incompatible phases must generically be first order, except at
multicritical points. Second, the formulation of the theory
of the phase transition in terms of fields other than the
order parameters of the proximate phases. At a technical level
this happens when the ordering is driven by dangerously
irrelevant operators.

As we have discussed in this paper, the situation near the RK
points is more complicated than was assumed in Ref.~\onlinecite{fn1}.
Instead of a single tilting transition between the plaquette
phase and the staggered phase there is a devil's staircase of commensurate
incommensurate transitions. Nevertheless, sufficiently close to
the RK point, any commensurate ordering is extremely weak and for
practical purposes one can think of the RKFPs controlling a
continuous transition between a flat phase with plaquette order
and a tilted phase with staggered order as well as incommensurate
order with a continuously varying wavevector. The weakly tilted phase
however is (logarithmically) deconfined, unlike the fully tilted
staggered phase.

In the quantum dimer models analyzed in this paper, the RKFPs control
multicritical surfaces---in contradiction with the claim in
Ref.~\onlinecite{fn1}. However this is not inevitable in this class of
problems. A quantum vertex model discussed in the appendix exhibits
a critical surface controlled by RKFPs. In either case the underlying
height action \pref{eq:heightact} is not derived from the order
parameters of the proximate phases. In this sense they lie outside
the Landau paradigm when applied to quantum phase transitions.

Finally, we note that in the $d=3+1$ cubic QDM, the tilting transition
remains continuous for weakly perturbed QDMs and the tilted phase
exhibits only staggered correlations. This is now a continuous
transition inside a deconfined region where Landau theory predicts a
continuous transition. That said, the critical theory is still not
what one would naively deduce from an order parameter analysis of
the ordered (staggered) phase---the correlations of the order parameter
at criticality have a dipolar form.
The complication is the need to properly treat the
local constraint $\nabla \cdot {\bf E} =0$, 
which is accounted for in the
action Eq.~\ref{eq:actiongauge} by the introduction of the vector
potential $\bA$. It is also the case that the ordered phase breaks
rotational symmetry due to a dangerously irrelevant operator. Altogether
while the three dimensional problem isn't quite as unusual as the
two dimensional one, it also furnishes an instance where the critical
theory requires a departure from the standard cookbook.

\section{Summary}
\label{sec:summary}

We have analyzed the phase diagram
in the
vicinity of the RK points in the bipartite QDMs. In $d=2$ we find
a first order transition separating confining plaquette phases from a
devil's staircase of commensurate/incommensurate and confining/deconfining
valence bond crystals on both the honeycomb and square lattices with
the RK points sitting on a multicritical surface of codimension two.
In $d=3$ the RKFP controls a continuous transition between an
RVB phase and a deconfined staggered valence bond crystal. QDMs
that exhibit these phenomena can be constructed by adding multi-dimer
potential energies to the RK point Hamiltonians. Whether the
phenomena here can be identified in a spin model or in other physical
systems, is an interesting topic for future work. Finally, we note
that as we were writing up this work there appeared
Ref.~\onlinecite{vishwanath}
which has considerable overlap with our work but which differs
from our conclusions on some points.

\begin{acknowledgments}
We would like to thank P. Fendley, C. Henley,
J. Kondev, S. Sachdev and T. Senthil for useful discussions; the
latter especially for focusing our attention on the implications
of lattice symmetries. This work was in
part supported by the Minist\`ere de la Recherche et des Nouvelles
Technologies with an ACI grant (RM), by the
National Science Foundation grants NSF-DMR-9978074 and NSF-DMR-0213706
at Princeton University (SLS, DAH and VO), NSF-DMR-01-32990 at the
University of Illinois (EF),  and by the David and Lucile
Packard Foundation (SLS and VO).
\end{acknowledgments}

\appendix
\section{Quantum Vertex Models}
\label{sec:vertex}

Quantum vertex models are close cousins of quantum dimer models. Here we begin
with a Hilbert space labelled by configurations of a classical vertex model,
i.e. each bond hosts an Ising variable which we picture as an arrow marked on
it. Then we introduce a local quantum dynamics which consists of reversing
local closed loops of arrows. In cartoon form, the resonance move analogous to
the one captured in Eq.~\ref{eq:Hqdm} is given by
\bea
\includegraphics[width=0.2\textwidth]{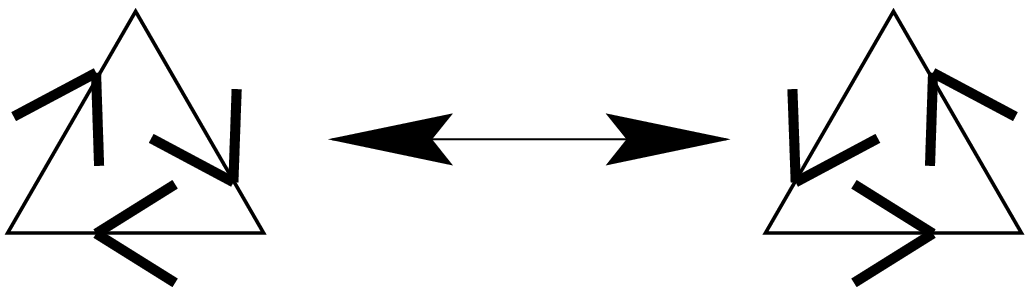}.
\eea
As the bond variables here are oriented their
physical origins will neccessarily be distinct from those of dimers. For
example, a quantum six vertex model was introduced in Ref.~\onlinecite{mts} as
a model for quantum effects in two dimensional ice as well as via a mapping
from a planar pyrchlore Ising magnet placed in a transverse magnetic field.
Subsequently it also been argued to arise as an effective theory of the
quantum fluctuations in the d-density wave state where the microscopic
variables are oriented currents.\cite{sc-vertex}

An exhaustive analysis of the RK point manifold in the quantum
eight vertex model is contained in Ref.~\onlinecite{aff}. 
When critical,
the model is equivalent to the six vertex model. In this case
the conservation law at the vertices leads straightforwardly
to a height representation in which the two sublattices of the
square lattice host even and odd values of a height field with
circumference two. A nice feature of the quantum six vertex problem
is that one can find a family of RK points at which the value of
$\rho_4$ can be tuned continuously; the ground state wavefunctions
in these cases are no longer equal amplitude superpositions but still
exhibit isotropic critical correlations. The analysis of perturbations
of these RK points now closely parallels our discussion of the
square lattice QDM and we conclude that they govern a multicritical
surface, likely between a plaquette phase \cite{iceplaquette} and a
devil's staircase region. The significant differences from the dimer
case involve a) the degeneracies of the phases and b) the varying
dimension of the vertex operator piece of the dimer operator. The
latter causes the relative strengths of the staggered order and
incommensurate order to vary in the tilted phase. For example, near
the equal amplitude RK point, the staggered order is quadratic in
the tilt as against cubic for the incommensurate piece.

Of the 32 possible vertices on a triangular lattice, 20 respect the
`ice rule' of having the same number of arrows pointing in an out.
The corresponding 20-vertex model (the ice model on the triangular
lattice) thus permits a height representation. The
heights live on the dual honeycomb lattice and are even and odd
on its two sublattices with compactification radius two. 
The symmetries
now force the height action to be even in $h$ and isotropic to
fourth order in gradients. Consequently the line of RK fixed
points is now stable with a dangerously irrelevant sixth order
term picking one of six lattice directions to orient the tilted
phase. The $(h, {\bf x})$ lattice is an hcp structure with
in-plane bonds deleted and its reciprocal lattice vectors
govern the commensurate states in the devil's staircase region.

The values of $\rho_4$ stemming from particular microscopic
models in this family are under investigation and will be
reported more fully elsewhere;\cite{im} here we content ourselves
with noting that at the equal amplitude RK point the height {\it is}
rough\cite{im} and hence in its neighborhood we now have an example of
critical behavior governed by RK fixed points and a Cantor deconfined
region which can be accessed without dialing the extra parameter needed
in the dimer models discussed previously in this paper. Based
on experience with this class of models, it seems likely that
the flat phase is again a plaquette phase which is symmetry incompatible
with the Cantor region.

\end{document}